\documentclass[9pt,a4paper,twoside]{tau-class/tau}
\usepackage[spanish]{babel}

\journalname{bor}
\title{ Gravitational DePhaser: A Temporal Gravitational Lensing Model for Creative Audio Processing}

\author[a]{Carlos Darío Badilla Cerdas }

\affil[a]{Escuela de Física, Universidad de Costa Rica}
\affil{\url{dari.badilla@gmail.com}}

\institution{Universidad de Costa Rica}
\footinfo{|borterion$\rangle$}

\institution{Universidad de Costa Rica}
\footinfo{borterion}

\begin{abstract}
Classical delay, chorus, and reverb effects model multiple propagation paths using discrete delays or statistical distributions. This work proposes a model inspired by gravitational lensing, where an audio signal is treated as a wave packet traversing a temporally distorted potential rather than physical spacetime. The distortion generates multiple effective propagation paths with different delays, phases, and gains, producing controlled constructive and destructive interference.
\end{abstract}


\begin{document}
		
    \maketitle 
    \thispagestyle{firststyle} \tauabstract 

\section{Introduction}

\begin{figure} [H]
	\centering
	\includegraphics[width=0.9\linewidth]{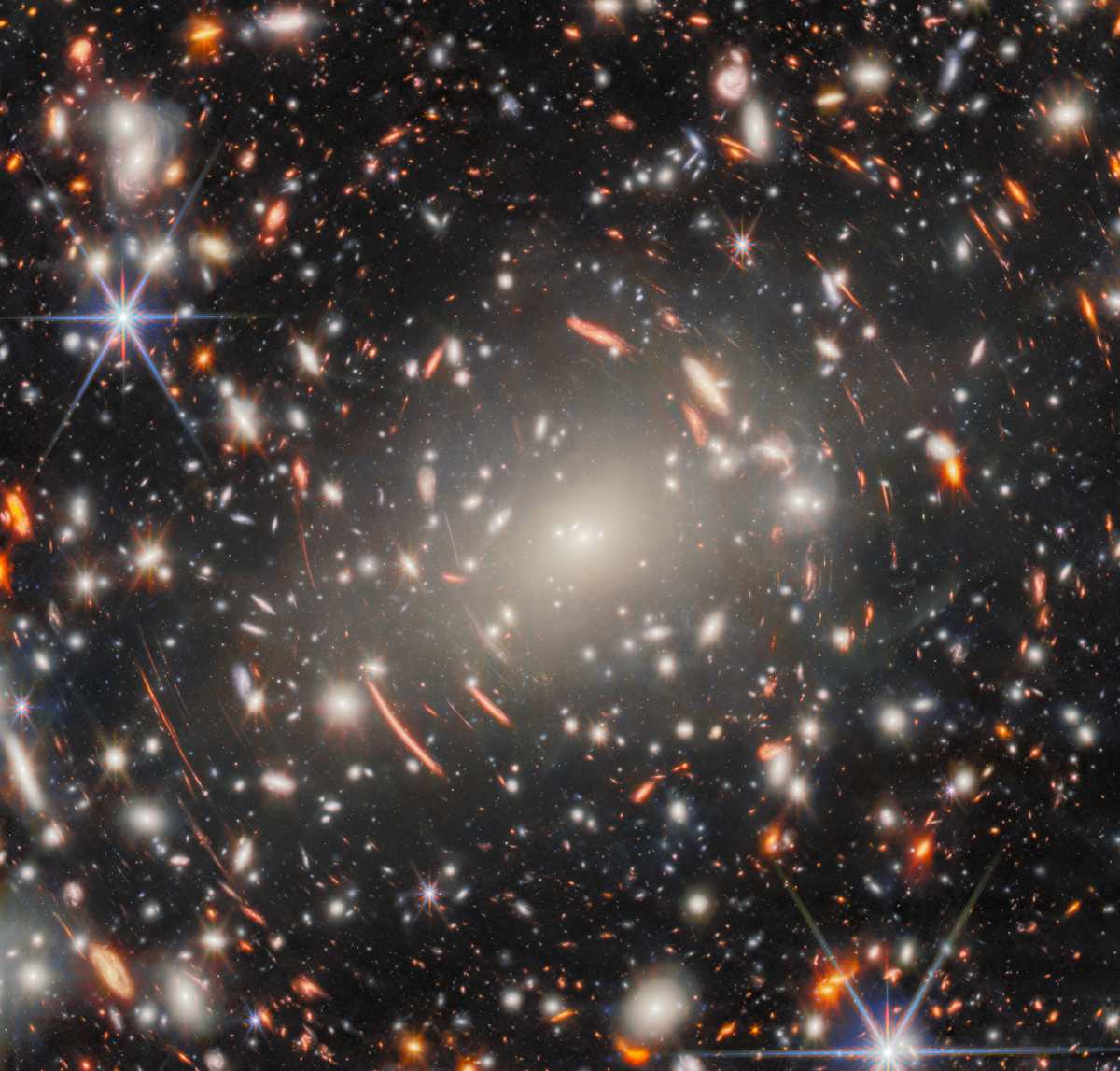}
	\caption{Gravitational lenses visible in the Abell S1063 Cluster. Photograph obtained by ESA/Webb, NASA and CSA, H. Atek, M. Zamani (ESA/Webb)- Acknowledgement: R. Endsley}
	\label{fig:webbglimpsesthedistantpast}
\end{figure}

The interaction between physics and sound synthesis has historically produced a wide range of computational tools, from physical modeling instruments to signal processing techniques inspired by wave propagation phenomena. In many cases, concepts originally developed to describe natural systems provide novel perspectives for the design of audio effects and musical interfaces.
One of the most remarkable consequences of General Relativity is gravitational lensing, a phenomenon in which the curvature of spacetime modifies the trajectories followed by propagating electromagnetic waves \cite{schneider2006gravitational}. Figure \ref{fig:webbglimpsesthedistantpast} shows an example of strong gravitational lensing in the cluster Abell S1063. The arcs curved around the bright center immediately catch the eye: they carry information about what lies behind the massive object, but the price paid — by the laws of the universe themselves — is distortion.

\begin{quoting}
	\textit{Rather than traveling along a single path, radiation may reach an observer through multiple effective trajectories, producing observable effects such as magnification, delay, and interference \cite{thorne2000gravitation}.}.
\end{quoting}

This work explores an analogous concept in the context of digital audio processing. Instead of modeling physical spacetime directly, we introduce an effective delay-generating potential that modifies the propagation of an audio signal through a set of virtual paths. Each path contributes a delayed and phase-shifted version of the original waveform, generating interference structures that can be controlled through parameters inspired by gravitational lens models.

The proposed framework is not intended as a physically accurate simulation of gravitational lensing. Instead, it borrows the mathematical language of lensing potentials as a generative paradigm for audio effect design. By interpreting delay lines as effective propagation trajectories shaped by a deformable potential, the model provides a unified approach capable of producing chorus, flanging, diffusion, and reverberation-like behaviors.

A real-time implementation was developed using the JUCE framework \cite{juce}, demonstrating how physically inspired propagation models can serve as practical tools for sound design while providing an accessible bridge between concepts from relativistic physics and computational audio.

\section{Mathematical Model}

The proposed audio effect is inspired by the phenomenon of gravitational lensing, where electromagnetic waves propagate through curved spacetime and may follow multiple effective trajectories before reaching an observer. Rather than simulating physical gravitational lensing directly, we define an effective delay-generating potential acting on an audio waveform. This potential plays a role analogous to a geometry — it assigns a propagation delay to each virtual path — without claiming the formal structure (metric, curvature tensor) of an actual spacetime.
Let

\begin{equation}
x(t)
\end{equation}

be the input audio signal.

A virtual gravitational mass $M$ generates a temporal distortion field that modifies the propagation time of the signal. The resulting output is modeled as the coherent superposition of $N$ effective propagation paths:

\begin{equation}
y(t)=\sum_{i=1}^{N} a_i,x!\left(t-\tau_i\right)e^{j\phi_i}, \label{delta}
\end{equation}

where $a_{i}$ represents the amplitude contribution of the $i-th$ path, $\tau_{i}$ is the associated temporal delay, and $\psi_{i}$ is a phase offset.

The delays are generated through an effective potential

\begin{equation}
\Psi(\theta)=M
\exp\left(
-\frac{(\theta-\mu)^2}{2\sigma^2}
\right),
\end{equation}

where $\theta$ denotes a virtual angular coordinate, $\mu$ defines the center of the lens, and $\sigma$ controls its spatial extent.
The delay associated with each trajectory is defined as

\begin{equation}
\tau_i=\tau_0+\alpha \Psi(\theta_i),
\end{equation}

where $\tau_0$ is a base propagation time and $\alpha$ controls the strength of the distortion.

The amplitude weighting is given by

\begin{equation}
a_i=
\frac{\Psi(\theta_i)}
{\sum_{k=1}^{N}\Psi(\theta_k)},  \label{eq5}
\end{equation}

ensuring energy normalization across all effective paths.

\subsection{Dynamic Curvature Modulation}

To generate chorus-like, flanger-like, and diffuse reverberant behaviors, the virtual mass parameter is allowed to evolve in time:

\begin{equation}
M(t)=M_0+\Delta M \sin(2\pi f_m t), \label{masad}
\end{equation}

where $f_m$ is the modulation frequency.

Consequently,

\begin{equation}
\tau_i(t)=
\tau_0+
\alpha
\Psi(\theta_i,t),
\end{equation}

creating continuously varying propagation paths.

This formulation produces constructive and destructive interference patterns analogous to those observed in wave propagation through gravitationally distorted media. Depending on the number of paths $N$ and the modulation parameters, the system can operate as a chorus, flanger, diffuser, or experimental reverberation processor.

\subsection{Effective Lensing Potential}

An alternative formulation employs a logarithmic potential inspired by gravitational lens models:

\begin{equation}
\Psi(\theta)=
M\ln\left(
1+\frac{\theta^2}{r_c^2}
\right),
\end{equation}

where $r_{c}$c is a virtual core radius controlling the concentration of the distortion field.
The corresponding delay distribution becomes

\begin{equation}
\tau_i=
\tau_0+
\alpha
M
\ln\left(
1+\frac{\theta_i^2}{r_c^2}
\right),
\end{equation}

providing broader delay distributions and richer interference structures suitable for ambient and spatial audio applications.

\section{Geometric Interpretation and Audio Processing Framework}

The proposed model can be interpreted as an effective propagation geometry acting on an audio waveform. In this framework, the input signal is treated as an incident wave propagating through a virtual distortion field generated by a mass parameter $M$.

Rather than following a single trajectory, the waveform propagates through multiple effective paths surrounding the virtual mass. Each path possesses a different geometric length and therefore a distinct propagation delay. When these delayed copies are recombined at the output, constructive and destructive interference patterns emerge.

\begin{figure}
	\centering
    \includegraphics[width=0.9\linewidth]{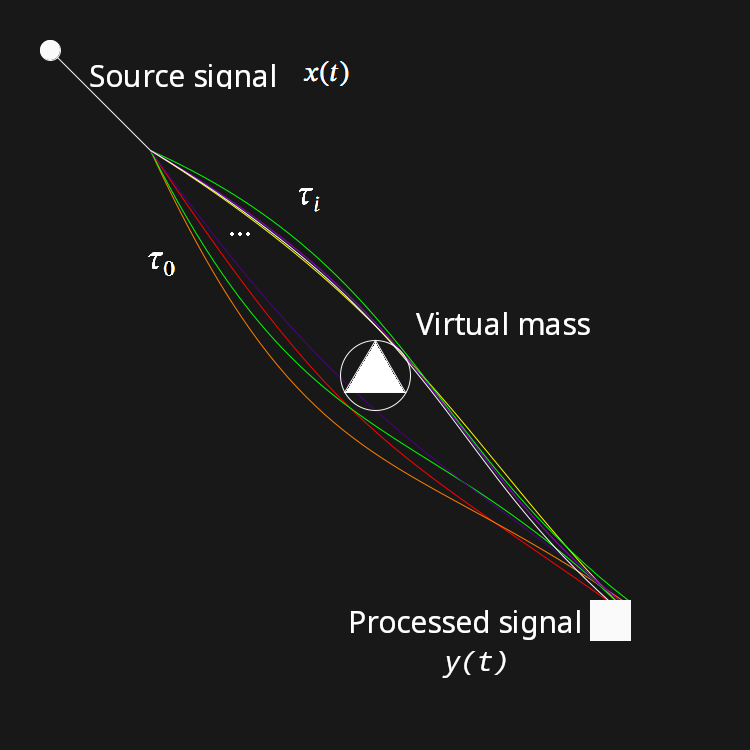}
    \caption{The dark side of the conceptual model of signal trajectories.}
    \label{fig:placeholder}
\end{figure}

Figure \ref{fig:placeholder} illustrates the conceptual model adopted in this work. Colored trajectories represent independent propagation paths connecting the sound source and the observer. The curvature of each path is controlled by the virtual mass parameter, which determines the distribution of arrival times at the output.
From a signal processing perspective, each trajectory corresponds to an individual delay line whose parameters are derived from the geometric model. The resulting superposition produces spectral and temporal modifications analogous to those observed in chorus, flanging, diffusion, and reverberation effects.
This interpretation establishes a direct conceptual bridge between gravitational lensing phenomena and creative audio processing, where path multiplicity and propagation delays become perceptually meaningful musical parameters.

\section{Implementation Considerations}

A real-time implementation of the proposed model was developed using the JUCE framework. The processing architecture consists of a set of parallel delay lines representing the effective propagation paths generated by the virtual lensing field.

For each processing block, delay values are computed from the geometric model and assigned to the corresponding propagation paths. The delayed signals are subsequently weighted and recombined according to Eq. (\ref{eq5}), producing the final output waveform.
s
The number of active paths, virtual mass, temporal spread, and path weighting functions can be modified in real time, allowing the system to transition continuously between subtle modulation effects and highly diffused spatial textures.

Since each of the N paths requires one delay-line read and one multiply-add per output sample, the computational cost of Eq. (\label{delta}) scales as $O(N)$ per sample. This linear scaling makes the approach suitable for real-time operation on contemporary audio hardware. Informal testing during development suggests the plugin runs without audible dropouts at typical buffer sizes on a conventional laptop DAW setup, though systematic benchmarking across host applications and buffer sizes has not yet been performed.

During the development of the initial versions, a bug was discovered - not in the code itself, but in the formulation of the system's constraints. For Eq. (\ref{masad}), if $\Delta M > M_{0}$, the modulation term can drive the effective mass $M(t)$ negative for part of its cycle, since $\sin(2 \pi f_{m}t)$ reaches $-1$. A negative effective mass fed into Eq. (\ref{delta}) leads to divergent results that rapidly approach infinity, producing output levels of $+100 dBRMS$.

Consequently, the boundary conditions had to be reconsidered, requiring that $M_{0} > \Delta M$. This resolved the issue by keeping $M(t)$ strictly positive for all $t$, avoiding the problematic range without compromising system performance.

\subsection{Performance}
The resulting VST3 plugin was tested on a Lenovo IdeaPad A10-9600P machine with 12 GB of RAM, alongside a Native Instruments Komplete Control 2 audio interface. The stability test was carried out in the Reaper DAW and consisted of one minute of audio playback at 48 kHz, with every parameter set to maximum level. The test was repeated with block sizes ranging from 4 up to 2048.

\begin{figure}[h]
	\centering
	\includegraphics[width=0.9\linewidth]{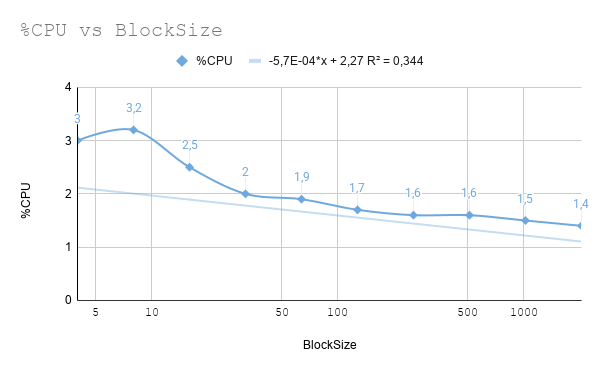}
	\caption{CPU percentage trend as a function of sample block size. An inverse relationship is observed, with demand decreasing as block size increases.}
	\label{fig:cpu-vs-blocksize}
\end{figure}

With the smallest block size, a single instance consumes 3\% of the CPU capacity.
Consumption decreases as the block size increases, reaching a minimum of 1.4\% at 2048.
In every case, consumption remains low, stable, and with a downward trend.

\section{Conclusions}

The proposed framework demonstrates how concepts originating from gravitational lensing can be translated into a practical signal processing paradigm for musical applications. While the model does not attempt to reproduce the full relativistic treatment of gravitational lensing, it preserves key observable features such as path multiplicity, propagation delay differences, and wave interference.
One of the primary advantages of the approach is the intuitive relationship between astrophysical concepts and audio parameters. Changes in virtual mass directly affect delay distributions, whereas the number of active paths regulates the density of interference structures created inside the system.
Finally, this framework suggests that simplified geometric interpretations of relativistic phenomena can provide both scientifically meaningful metaphors and practical tools for creative audio processing.

\subsection{Bye.}

\begin{quote}
	Space curves—\\
	your gaze tracing \\
	the thousand paths where \\
	I will not see you.
\end{quote}


\addcontentsline{none}{section}{References}
\printbibliography


\end{document}